\documentclass{icrc29}
\usepackage{graphicx,amssymb,amsmath,times}
\begin{document}
%Title of paper
\def\aprle{\buildrel < \over {_{\sim}}}
\def\aprge{\buildrel > \over {_{\sim}}}
\def\ee{e^{+}e^{-}}
\def\ll{\langle}
\def\rr{\rangle}
\def\ra{\rightarrow}
\def\lra{\leftrightarrow}
\newcommand{\nue}{\nu_{e}}
\newcommand{\num}{\nu_{\mu}}
\newcommand{\nut}{\nu_{\tau}}
\newcommand{\nus}{\nu_{sterile}}
\title[Search for exotic contributions ...]{Search for exotic contributions to atmospheric neutrino oscillations}
\author[G. Battistoni et al.] {G. Battistoni$^a$, Y. Becherini$^b$, S. Cecchini$^{b,c}$, M. Cozzi $^b$, H. Dekhissi$^{b,d}$, L. Esposito$^b$,
        \newauthor
	  G. Giacomelli$^b$, M. Giorgini$^b$,	G. Mandrioli$^b$, S. Manzoor$^{b,e}$,	A. Margiotta$^b$,	L. Patrizii$^b$,
	\newauthor
		V. Popa$^b,f$,	M. Sioli$^b$,	G. Sirri$^b$,	M. Spurio$^b$ and V. Togo$^b$
	\\
(a) INFN Sez. di Milano, 20133 Milano, Italy \\
(b) Dip. Fisica dell'Universita' di Bologna and INFN, 40127 Bologna, Italy  \\
(c) INAF/IASF Sez. di Bologna, 40129 Bologna, Italy \\
(d) LPTP, Faculty of Sciences, University Mohamed 1st, B.P. 424, Oujda, Morocco \\
(e) PRD, PINSTECH, P.O. Nilore, Islamabad, Pakistan  \\
(f) ISS, 77125 Bucharest-Magurele, Romania}
\presenter{Presenter: A. Margiotta (margiotta@bo.infn.it), \  
ita-margiotta-A-abs2-he14-poster}
\maketitle

\begin{abstract}
The energy spectrum of neutrino-induced upward-going muons in MACRO was analysed
in terms of relativity principles violating effects, keeping standard
mass-induced atmospheric neutrino oscillations as the dominant source of
$\num \ra \nut$ transitions. The data disfavor these 
possibilities even at a sub-dominant level; stringent 90\% C.L.
limits are placed on the Lorentz invariance violation parameter
$|\Delta v| < 6 \times 10^{-24}$ at $\sin 2{\theta}_v$ = 0 and
$|\Delta v| < 2.5 \div 5 \times 10^{-26}$ at $\sin 2{\theta}_v$ = $\pm$1.
The limits can be re-interpreted as bounds on the Equivalence Principle  violation
parameters.
\end{abstract}

\section{Introduction}
The phenomenon of neutrino flavor oscillations, induced by flavor-mass eigenstate mixing, is considered the favored solution for solar and atmospheric
neutrino data \cite{solar}-\cite{soudan2} over a wide range of alternative solutions \cite{macro-sterile}-\cite{eclipse}. These latter mechanisms were considered under the hypothesis that each one of them solely accounts for the observed effects. 
Here we address the possibility of a mixed scenario: one mechanism,
the mass-induced flavor oscillations, is dominant and a second
mechanism is included as sub-dominant: it could be neutrino flavor transitions induced by violations of relativity principles, i.e. violation of the Lorentz invariance (VLI) or of the equivalence principle (VEP). In this mixed scenario, assuming that neutrinos can be described in terms of three
distinct bases - flavor, mass and velocity eigenstates -
the latter being characterized by different maximum attainable velocities (MAVs), and
by considering that only two families contribute to the atmospheric $\nu$ violation oscillations, the $\nu_\mu$ survival probability can be expressed as~\cite{fogli}-\cite{glashow04}
\begin{equation}
P_{\nu_\mu \to \nu_\mu} = 1 - \sin^2 2 \Theta \sin^2 \Omega
\label{eq:due}
\end{equation}
where the global mixing angle $\Theta$ and the term $\Omega$ are given by:
\begin{equation}
\begin{array}{ll}
2\Theta = \arctan (a_1/a_2) , ~\Omega = \sqrt{\left( a_1^2 + a_2^2 \right)}~.
\end{array}
\label{eq:tre}
\end{equation}
The terms $a_1$ and $a_2$ in Eq.~(\ref{eq:tre}) contain the relevant physical information
\begin{equation}
\begin{array}{ll}
a_1 = 1.27 | \Delta m^2 \sin 2 \theta_m L/E + 2 \cdot 10^{18} \Delta v \sin 2 \theta_v
LE e^{i \eta} | \\
a_2 = 1.27\left( \Delta m^2 \cos 2\theta_m L/E + 2 \cdot 10^{18} \Delta v \cos 2
\theta_v LE \right) ~,
\end{array}
\label{eq:quattro}
\end{equation}
where the muon neutrino pathlength $L$ is expressed in km, the neutrino
energy $E$ in GeV and the oscillation parameters $\Delta m^2 = m^2_{\nu_3^m} -
m^2_{\nu_2^m}$ and $\Delta v = v_{\nu_3^v} - v_{\nu_2^v}$ are in eV$^2$ and
$c$ units, respectively. The unconstrained phase $\eta$ refers to the connection
between mass and velocity eigenstates.
The whole domain of variability of the parameters can be accessed with the requirements
$\Delta m^{2} \ge 0$, $0 \le \theta_m \le \pi/2$, $\Delta v \ge 0$
and $-\pi/4 \le \theta_v \le \pi/4$.\\
The same formalism also applies to violation of the equivalence principle,
after substituting $\Delta v/2$ with the adimensional product $|\phi| \Delta \gamma$;
$\Delta \gamma$ is the difference of the coupling constants for neutrinos
of different types to the gravitational potential $\phi$~\cite{gasperini}.\\
\begin{figure}[h]
\centering
%\hspace{-.4cm}
\begin{minipage}[c]{0.6\textwidth}
\centering
\hspace{-0.5in}
\includegraphics*[width=0.5\textwidth,angle=0,clip]{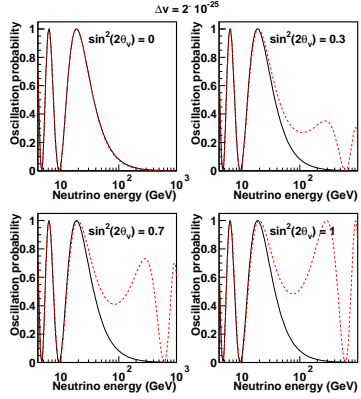}
\end{minipage}%
\begin{minipage}[c]{0.4\textwidth}
\centering
\caption{\label{fig:ftest}Energy dependence of the $\nu_\mu \rightarrow \nu_\tau$ oscillation probability
for mass induced oscillations alone (continuous curves), and mass-induced + VLI
oscillations for $\Delta v = 2 \times 10^{-25}$ and different values of the
$\sin^2 2{\theta}_v$ parameter (dashed curves). The
neutrino pathlength was fixed at $L = 10^4$ km.}
\end{minipage}%
\end{figure}

As shown in~\cite{glashow9799,glashow04,strumia},
the most sensitive tests of VLI can be made by analysing the high energy tail
of atmospheric neutrinos at large pathlengths values. Fig.~\ref{fig:ftest} shows the energy dependence of the $\num \ra \nut$ oscillation
probability as a function of the neutrino energy, for neutrino mass-induced
oscillations alone and for both mass and VLI-induced oscillations. Notice that for $m_{\nu} <$ 1 eV and $E_\nu >$ 100 GeV: $\gamma_L >$ 10$^{11}$.

\section{Exotic contributions to mass-induced oscillations}
In order to analyse the MACRO data in terms of VLI, we used the  
subsample of 300 upward-throughgoing muons whose energies were estimated via
Multiple Coulomb Scattering \cite{mcs1}-\cite{mcs2}.
Two independent and complementary analyses have been carried out: one based on the $\chi^2$ 
criterion and the Feldman and Cousins prescription~\cite{felcou}, and a second 
one based on the maximum likelihood technique~\cite{macro-vli}.\\

\vspace{-0.3in}
\subsection{$\mathbf{\chi^2}$ Analysis}
\label{chi} 
Following the analysis in~\cite{mcs2}, we selected a low and a high energy sample
by requiring that the reconstructed neutrino energy $E^{rec}_\nu$ should be
$E^{rec}_\nu <$ 30 GeV and $E^{rec}_\nu >$ 130 GeV.
The number of events surviving these cuts is $N_{low}$ = 49 and $N_{high}$ =
58; their median energies, estimated via Monte Carlo,
are 13 GeV and 204 GeV (assuming mass-induced oscillations).
We then keep the neutrino mass oscillation parameters fixed at the values of ~\cite{macro-last}. The factor $e^{i\eta}$  is assumed to be real ($\eta$ = 0 or $\pi$).
Then, we scanned the plane of the 2 free parameters
($\Delta v$, $\theta_v$) minimizing the $\chi^2$ function
comprehensive of statistical and systematic uncertainties~\cite{macro-vli}.\\
For the Monte Carlo simulation described in~\cite{mcs2} the 
neutrino fluxes in input is given by~\cite{newhonda}.
The largest relative difference of the extreme values of the MC expected
ratio $N_{low}/N_{high}$ is 13\%.
However, in the evaluation of the systematic error, the main sources
of uncertainties for this ratio
(namely the primary cosmic ray spectral index and neutrino cross sections)
have been separately estimated and their effects added in quadrature
(see~\cite{mcs2} for details):
in this work, we use a conservative 16\% theoretical systematic error
on the ratio $N_{low}/N_{high}$.The experimental systematic error on the ratio was estimated to be 6\%.
The inclusion of the VLI effect does not improve the $\chi^2$ in any point
of the ($\Delta v$, $\theta_v$) plane, compared to mass-induced oscillations
stand-alone, and limits on VLI parameters were obtained.
The 90\% C.L. limits on $\Delta v$ and
$\theta_v$, computed with the Feldman and Cousins prescription~\cite{felcou},
are shown by the dashed line in Fig.~\ref{fig:super}-left.
The energy cuts described above
were optimized for mass-induced neutrino oscillations. In order to
maximize the sensitivity of the analysis for VLI induced oscillations,
we performed a blind analysis, based only on Monte Carlo events, to
determine the energy cuts which yield the best performances. The results of this study
suggest the cuts $E^{rec}_\nu <$ 28 GeV and $E^{rec}_\nu >$ 142 GeV;
with these cuts the number of events in the real data are
$N^{\prime}_{low}$ = 44 events and $N^{\prime}_{high}$ = 35 events.
The limits obtained with this selection are shown in
Fig.~\ref{fig:super}-left by the continuous line.
As expected, the limits are now more stringent than for the previous choice.
In order to understand the dependence of this result with respect to the
choice of the $\overline{\Delta m^2}$ parameter, we varied them around the best-fit point. We found that a variation
of $\overline{\Delta m^2}$ of $\pm$30\% moves up/down the upper limit of VLI
parameters by at most a factor 2.
Finally, we computed the limit on $\Delta v$ marginalized with respect to
all the other parameters left free to change inside the intervals:
$\Delta m^2$ = $\overline{\Delta m^2} \pm$30\%,
$\theta_m$ = $\overline{\theta}_m$ $\pm$20\%, $-\pi/4 \le \theta_v \le \pi/4$
and any value of the phase $\eta$. We obtained
the 90\% C.L. upper limit $|\Delta v| < 3 \times 10^{-25}$.
\vspace{-0.3in}
\subsection{Maximum Likelihood Analysis (MLT)}
\begin{figure}[h]
\vspace{-.4in}
\centering
%\hspace{-.4cm}
\begin{minipage}[c]{0.5\textwidth}
\centering
\hspace{-.4cm}
\includegraphics*[width=0.8\textwidth,angle=0,clip]{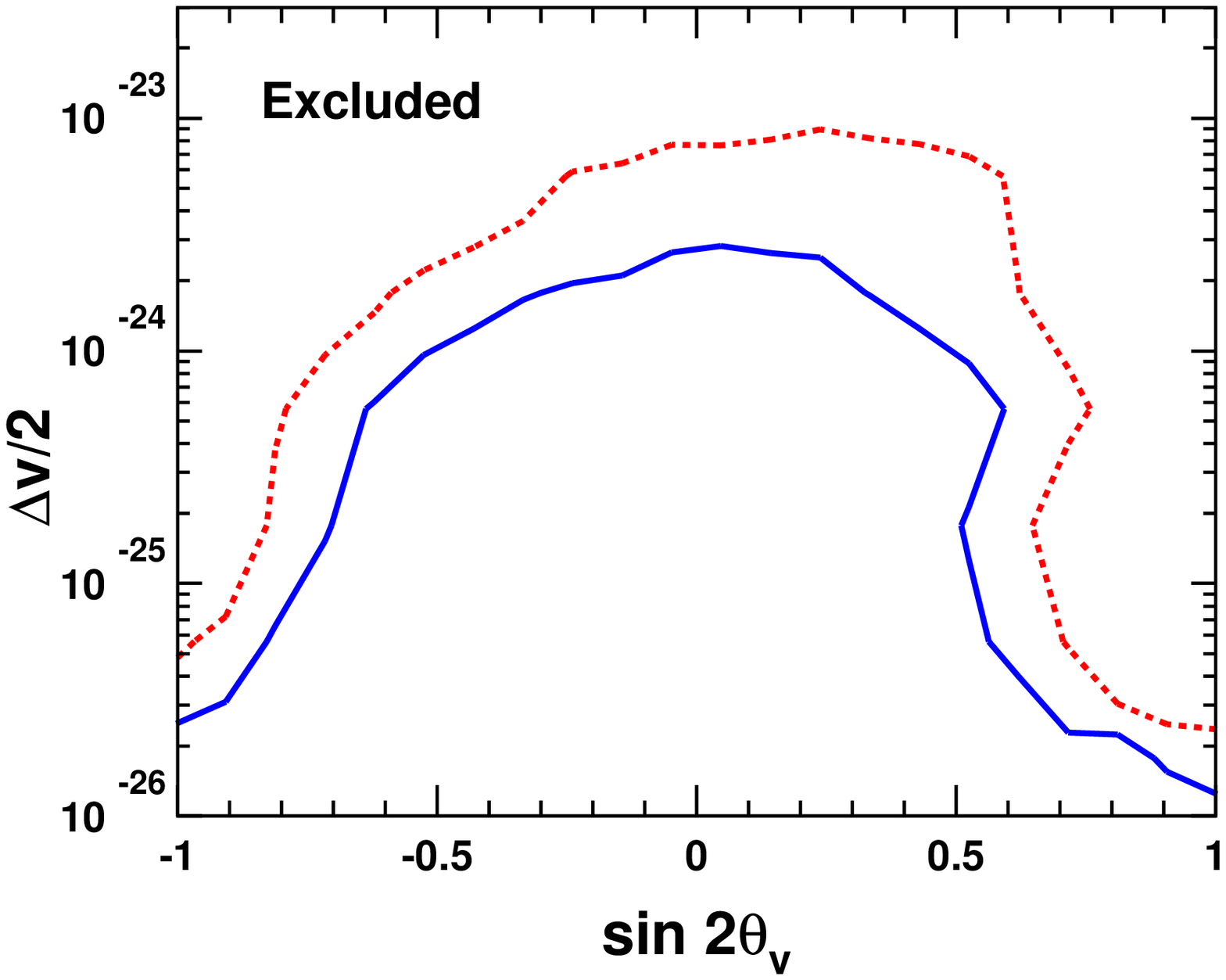}
\end{minipage}%
\begin{minipage}[c]{0.5\textwidth}
\centering
\includegraphics*[width=0.6\textwidth,angle=0,clip]{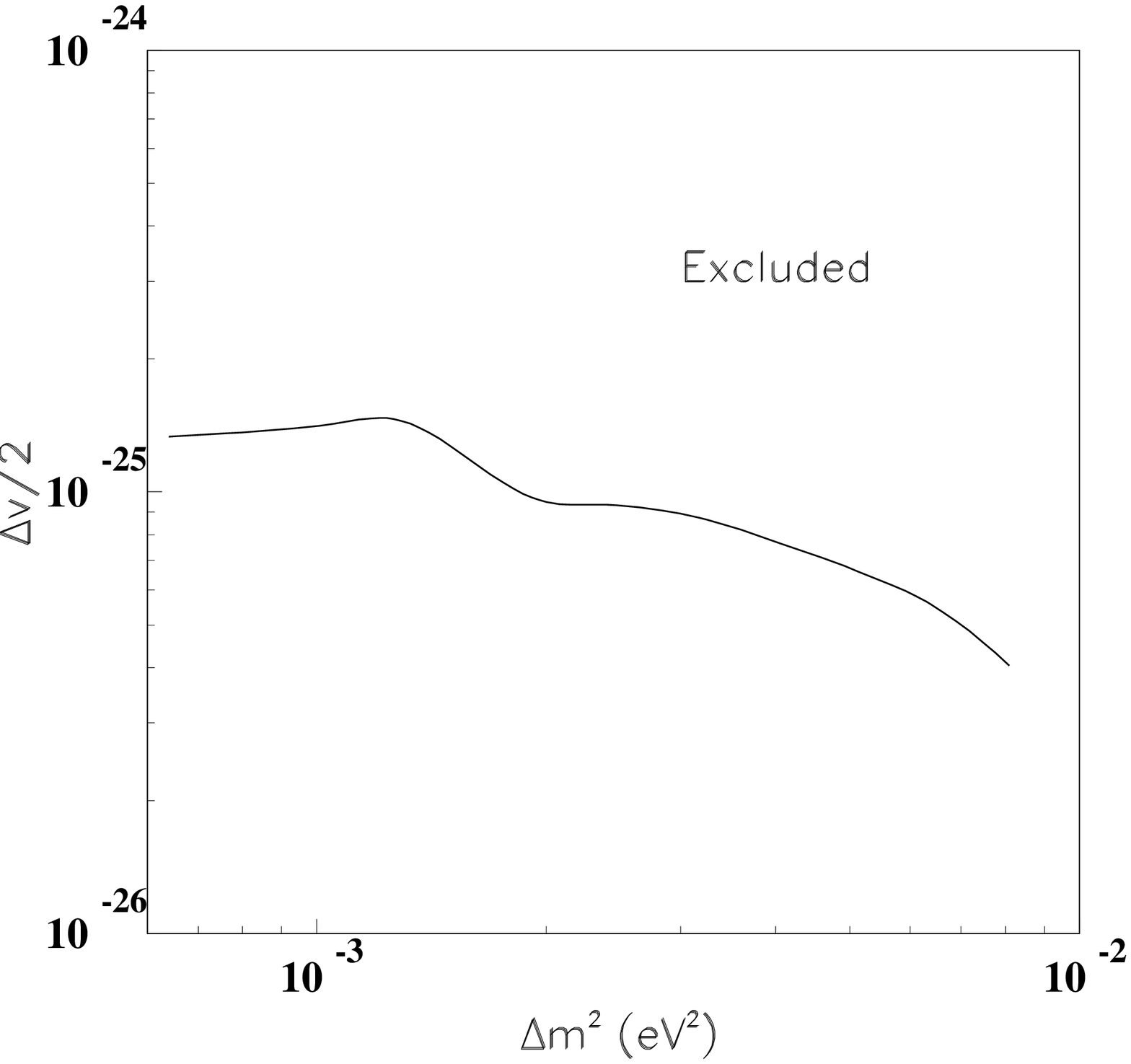}
\end{minipage}%
\vspace{-0.3in}
\caption{\label{fig:super}-$\bold{left}$: 90\% C.L. upper limits on the Lorentz invariance violation parameter $\Delta v$ versus sin $2\theta_v$.
  Mass induced oscillations are assumed in
  the two-flavor $\num \ra \nut$ approximation,
  with $\Delta m^2$ = 0.0023 eV$^2$ and $\theta_m$ = $\pi/4$.
  The dashed line shows the limit obtained with the same selection criteria of
  Ref.~\protect\cite{mcs2} to define the low and high energy samples; the
  continuous line is the result obtained with the selection criteria
  optimized for the present analysis (see text).
  \label{fig:vlad}-$\bold{right}$:
  90\% CL upper limits on the $\Delta v/2$  parameter versus
$\Delta m^2$ varying inside the 90\% CL MACRO global result~\protect\cite{macro-last}.}
\end{figure}

%\label{like}
A different and complementary analysis of VLI contributions to the atmospheric
neutrino oscillations was made on the MACRO muon data corresponding to
 parent neutrino energies in the range 25 GeV $\leq E \leq$ 75 GeV. This energy
region is characterized by the best energy reconstruction, and the number of muons
satisfying this selection is 106. These events are outside the energy ranges used
in the analysis discussed in Section 2.1, and thus the expected sensitivity
to VLI (or VEP) contributions to the atmospheric neutrino oscillations should be
lower; on the other hand, the maximum likelihood technique (MLT) has the advantage
to exploit the information event-by-event (it is a bin-free approach).
Given a specific hypothesis, MLT allows to determine the set of parameters 
$\mathbf{a} = (\Delta m^2, \theta_m, \Delta v, \theta_v)$)
that maximizes the probability of the realization of the actual measurements 
$\mathbf{x} = (E,L)$; it was accomplished by minimizing  
negative a log-likelihood function $\mathcal{L}$ \cite{rpp}. 
We have chosen different fixed values of the
$\Delta m^2$ and $\sin^2 2 \theta_m$ mass-oscillation parameters in~\cite{macro-last} 
and found the relative $\Delta v$ and $\sin^2 2 \theta_v$ that maximize the $f(\mathbf{x}_i;\mathbf{a})$ function proportional to the probability of realization of a given event.
%Eq.~(\ref{eq:log}). 
Fig. \ref{fig:vlad}
shows the 90\% CL upper of the VLI parameter $\Delta v/2$ versus the
assumed $\Delta m^2$ values.
\vspace{-0.1in}
\section{Conclusions}
 We have searched for ``exotic'' contributions to standard mass-induced
atmospheric neutrino oscillations arising from a possible violation of Lorentz
invariance using two different and complementary analyses. The first approach uses two sub-sets of events
referred to as the low energy and the high energy samples. The mass neutrino
oscillation parameters have the values of~\cite{macro-last}, 
and we mapped the evolution of the $\chi^2$ estimator
in the plane of the VLI parameters, $\Delta v$ and $\sin^2 2 \theta_v$. No
$\chi^2$ improvement was found, so we applied the Feldman Cousins method
to determine 90\% CL limits on the VLI parameters:
$|\Delta v| < 6 \times 10^{-24}$ at $\sin 2{\theta}_v$ = 0 and
$|\Delta v| < 2.5 \div 5 \times 10^{-26}$ at $\sin 2{\theta}_v$ = $\pm$1.
In terms of the parameter $\Delta v$ alone (marginalization with respect
to all the other parameters), the VLI parameter bound is (at 90\% C.L.)
$|\Delta v| < 3 \times 10^{-25}$.
These results may be reinterpreted in terms of 90\% C.L. limits of
parameters connected with violation of the equivalence principle,
giving the limit $|\phi \Delta \gamma| < 1.5 \times 10^{-25}$.
The second approach exploits the information contained in a data sub-set
characterized by intermediate muon energies. It is based on the MLT, and 
considers the mass neutrino oscillation parameters inside the 90\%
border of the global result~\cite{macro-last}.
The obtained 90\% CL limit on the $\Delta v$ VLI parameter
is also around $10^{-25}$.
\vspace{-0.1in}
\section{Acknowledgements}
We acknowledge the cooperation of the members of MACRO. We
thank many colleagues for discussions and advise, in particular
B. Barish, P. Bernardini,  A. De R\'{u}jula,
G.L. Fogli,  S.L. Glashow, P. Lipari and F. Ronga.


\vspace{-0.1in}
\begin{thebibliography}{99}

\bibitem{solar} 
J.~N.~Bahcall et al., JHEP, 408, 16 (2004).

\bibitem{macro-98} 
M.~Ambrosio et al., Phys. Lett., B434, 451 (1998).

\bibitem{macro-last}
M.~Ambrosio et al., Eur. Phys. J., C36, 323 (2004).

\bibitem{sk-general} 
Y.~Fukuda et al., Phys. Rev. Lett., 81, 1562 (1998).

Y.~Ashie et al., hep-ex/0501064.

\bibitem{soudan2}
M.~Sanchez et al., Phys. Rev., D68, 113004 (2003).

\bibitem{macro-sterile}
M.~Ambrosio et al., Phys. Lett., B517, 59 (2001).

\bibitem{sk-sterile} 
S.~Fukuda et al., Phys. Rev. Lett., 85, 3999 (2000).

\bibitem{sk-habig} 
A.~Habig, Proc 28th ICRC, (Tsukuba,  2003) Conf. Papers, 1, 1255.

\bibitem{vlad-oujda} V.~Popa and M. Rujoiu, in {\it CR: from Astronomy to Particle Physics},
%[G. Giacomelli et al. eds.]
(Dordrecht, 2001) 181.

\bibitem{fogli} 
G.~L.~Fogli et al., Phys. Rev., D60, 053006 (1999).

M.~C.~Gonzalez-Garcia and M.~Maltoni, Phys. Rev., D70, 033010 (2004).

\bibitem{glashow9799} 
S.~R.~Coleman and S.~L.~Glashow, Phys. Lett. B405, 249 (1997); Phys. Rev., D59, 116008 (1999).

\bibitem{glashow04} 
S.~L.~Glashow, hep-ph/0407087.

\bibitem{sk-dip} 
Y.~Ashie et al., Phys. Rev. Lett., 93, 101801 (2004).

\bibitem{eclipse} 
S.~Cecchini et al., Astropart. Phys., 21, 25 (2004); 21, 183 (2004).

\bibitem{gasperini} 
M.~Gasperini et al., Phys. Rev., D38, 2635 (1988).

\bibitem{strumia} 
A.~de Gouvea et al., Nucl. Phys., B623, 395 (2002).

\bibitem{mcs1} 
M.~Ambrosio et al., Nucl. Instrum. Meth., A492, 376 (2002).

\bibitem{mcs2} 
M.~Ambrosio et al., Phys. Lett., B566, 35 (2003).

\bibitem{felcou} 
G.~J.~Feldman and R.~D.~Cousins, Phys. Rev., D57, 3873 (1998).

\bibitem{macro-vli} 
G. Battistoni et al., Phys. Lett. B615, 14 (2005).

\bibitem{newhonda} 
M.~Honda et al., Phys. Rev., D70, 043008 (2004).

\bibitem{rpp}
S. Eidelman et al. (Particle Data Group), Phys. Lett. B592, 1 (2004).

\end{thebibliography}
\end{document}